\documentclass{ws-procs9x6}

\def\etal{{\it et al}}
\def\AEF{A.E. Faraggi}
\def\NPB#1#2#3{Nucl. Phys. B \textbf{#1},  (#3) #2}
\def\PLB#1#2#3{Phys. Lett. B \textbf{#1},  (#3) #2}
\def\PLA#1#2#3{Phys. Lett. A \textbf{#1},  (#3) #2}
\def\PRD#1#2#3{Phys. Rev. D  \textbf{#1},  (#3) #2}

\def\PRL#1#2#3{Phys. Rev. Lett. \textbf{#1},  (#3) #2}
\def\PRT#1#2#3{Phys. Rep. {\textbf#1},  #3 (#2)}
\def\MODA#1#2#3{Mod. Phys. Lett. A {\textbf #1},  (#3) #2}
\def\IJMP#1#2#3{Int. J. Mod. Phys. A {\textbf #1}, (#3) #2}

\begin{document}

\title{
\rightline{LTH--661}
\rightline{\tt hep-th/0509054}
\rightline{August 2005}
\rightline{}
Fictitious Extra Dimensions
}



\author{Alon E. Faraggi\footnote{to appear in the proceedings of
the fourth International Conference on Quantum Theory and Symmetries,
15--21/8/ 2005, Varna Bulgaria}
}

\address{Mathematical Sciences, University of Liverpool, Liverpool L69 7Zl\\
and~
Theoretical Physics, University of Oxford, Oxford, OX1 3NP}


\maketitle

\abstracts{
    String theory requires additional degrees of freedom to maintain
world--sheet reparameterisation invariance at the quantum level. These are
often interpreted as extra dimensions, beyond the 4 space-time. I
discuss a class of quasi-realistic string models in which
all the untwisted geometrical moduli are projected
out by GSO projections.
In these models the extra dimensions are fictitious, and
do not correspond to physical dimensions in a low energy
effective field theory. This raises the possibility that extra
dimensions are fictitious in phenomenologically viable string vacua.
I propose that self-duality
in the gravitational quantum phase--space provides
the criteria for the string vacuum selection.
}

\section{Introduction}
String theory, and its various modern incarnations, provides a consistent
and most developed framework to study the unification of all the observed
fundamental forces and interactions. This quest for unification
is an everlasting theme in modern physics. Early proponents
included Newton who unified celestial and terrestrial gravity;
Maxwell who unified the electric and magnetic forces; and 
Einstein who unified electromagnetism and mechanics. 
In more recent times all the observed fundamental processes in nature
are described in terms of the electromagnetic, weak and strong, gauge
interactions; and in terms of gravitational general relativity.
String theory affords the inclusion of all of those in a consistent
framework, and is the reason for its continued appeal and interest.
This, however, is not a speedy enterprise. Adjudicating whether it
succeeds or fails will likely require the efforts of more than one
generation. One should consider, however, that it took more than
two millenia to reach a decisive conclusion on heliocentrism versus
geocentrism. The reason being not merely the dogma
of well fashioned clergy, but rather the mundane interpretation
of the available data.

In the classical string we can
always gauge fix the two dimensional world--sheet metric
to the flat metric. Preserving this property in the
quantised string requires that we embed it in 26 space--time
dimensions in the case of the bosonic string; and 10 in the
case of the fermionic string. The closed string allows 
for independent treatment of the left-- and right--moving
modes on the string world--sheet. Hence, it gives rise to
the heterotic--string in which the left--movers
are fermionic and the right--movers are bosonic.

In the real world, we only observe four space--time dimensions, 
and internal symmetries of the particle spectrum.
The standard lore to rectify this apparent discrepancy is to
compactify the quantized string on an internal compactified manifold. 
In the case of the heterotic string 16 of the right--moving dimensions
are compactified on an even self--dual lattice
with fixed radii. Six right--moving coordinates, combined 
with six left--moving dimensions, are compactified on a six
dimensional real manifold, or on a three dimensional complex 
manifold. The size and shape of this internal compact manifold
are parametrized by the moduli. At present there is no known 
mechanism that selects and fix these moduli. Unravelling
it is one of the major hurdles facing string theory.

On the other hand, over the past two decades, phenomenological 
studies of string theory have continued in earnest, and numerous
quasi--realistic string models have been constructed. A natural 
question to ask therefore is whether these phenomenological string
vacua can offer a guide to the issue of moduli selection and fixation.
In this note I propose that the answer is affirmative. The
quasi--realistic heterotic string models in the
free fermionic formulation \cite{ffm}, which are associated with
$Z_2\times Z_2$ orbifold compactifications at special points
in the moduli space, points in the direction of the self--dual
point under T--duality as playing a special role in the 
vacuum selection, and to the independence of the left--right
moving modes as allowing for asymmetric conditions, that
result in fixation of all of the geometrical moduli, 
as well as all of the twisted sector moduli \cite{modfix}

\section{Moduli fixing in realistic string models}\label{rsm}

The general structure of the quasi--realistic 
free fermionic models and their phenomenological characteristics have
been amply discussed and reviewed in the past \cite{ffm}. 
Here I focus on the question of moduli fixing in these models.
The relation of these models to $Z_2\times Z_2$ orbifold
compactifications is elaborated in  \cite{z2z2}.
The untwisted sector of the $Z_2\times Z_2$ orbifold gives
rise to an $SO(10)$ GUT gauge group, which is broken down further,
by the string boundary conditions, to one of its sub--group.
The three twisted
sectors produce three spinorial 16 representations of $SO(10)$
decomposed under the unbroken $SO(10)$ subgroup.
In this manner the models give rise to three generations,
which possess the canonical $SO(10)$ GUT embedding. 
These models were primarily studied using the free
fermionic formalism \cite{fff},
in which all the string boundary conditions are given 
in terms of the free fermion transformation properties
on the string world--sheet. These fermionic models
correspond to bosonic compactifications, in which the moduli
are a priori fixed at a special point in the moduli space. 

The geometrical moduli are the untwisted
K\"ahler and complex structure moduli of the six dimensional
compactified manifold. Additionally, the string vacua contain
the dilaton moduli whose VEV governs the strength of the
four dimensional interactions. 
The VEV of the dilaton moduli
is a continuous parameter from the point of view of the perturbative
heterotic string, and its stabilization requires some nonperturbative
dynamics, or some input from the underlying quantum M--theory,
which is not presently available. 
The problem of dilaton stabilization is therefore not addressed
in this work, as the discussion here is confined to perturbative
heterotic string vacua. Additionally, the models contain 
twisted sector moduli. Since the moduli fields
correspond to scalar fields in the massless string spectrum,
the moduli space is determined by the set of boundary condition
basis vectors that define the string vacuum and encodes its
properties. The first step therefore is to identify the fields
in the fermionic models that correspond to the untwisted moduli.
The subsequent steps entail examining which moduli fields survive
successive GSO projections and consequently the residual moduli
space.  

The four dimensional fermionic heterotic string models are
described in terms of two dimensional conformal and superconformal
field theories of central charges $C_R=22$ and $C_L=9$, respectively.
In the fermionic formulation these are represented in terms
of world--sheet fermions. A convenient starting point
to formulate such a fermionic vacuum is a model
in which all the fermions are free. The free fermionic
formalism facilitates the solution of the conformal and
modular invariance constraints in terms of simple rules \cite{fff}.
Such a free fermionic model corresponds to a string vacuum
at a fixed point in the moduli space. Deformations from this
fixed point are then incorporated by including world-sheet
Thirring interactions among the world--sheet fermions,
that are compatible with the conformal and modular invariance
constraints. The coefficients of the allowed world--sheet Thirring
interactions correspond to the untwisted moduli fields.
For symmetric orbifold models, the exactly marginal operators
associated with the untwisted moduli fields take the general form
$\partial X^I{\bar\partial} X^J$, where $X^I$, $I=1,\cdots,6$, are
the coordinates of the six--torus $T^6$. Therefore, the untwisted
moduli fields in such models admit the geometrical interpretation of
background fields, which appear as couplings of
the exactly marginal operators in the non--linear sigma model action.
The untwisted moduli scalar fields are the background fields that are 
compatible with the orbifold point group symmetry. 

It is noted that in the Frenkel--Kac--Segal construction
of the Kac--Moody current algebra from chiral bosons,
the operator $i\partial X^I$ is a $U(1)$ generator of the 
Cartan sub--algebra. Therefore, in the fermionic
formalism the exactly marginal operators are given by Abelian
Thirring operators of the form $J_L^i(z){\bar J}_R^j({\bar z})$,
where $J_L^i(z)$, ${\bar J}_R^j({\bar z})$
are some left-- and right--moving $U(1)$ chiral currents
described by world--sheet fermions.
Abelian Thirring interactions preserve conformal invariance,
and are therefore marginal operators. One can therefore
use the Abelian Thirring interactions to identify the
untwisted moduli in the free fermionic models. The
untwisted moduli correspond to the Abelian Thirring
interactions that are compatible with the GSO projections
induced by the boundary condition basis vectors, which
define the string models. 

I now turn to examine the moduli space in concrete free fermionic
constructions. The models are constructed recursively
by adding additional boundary condition basis vectors,
which imposes GSO projections, truncating the existing spectrum,
and adding new sectors and new states.
The maximal moduli space of the $N=4$ vacuum at the free fermionic
point is the coset space $SO(6,22)/(SO(6)\times SO(22))$. Applying the
$Z_2\times Z_2$ projections truncates the untwisted moduli space to
$SO(2,2)/(SO(2)\times SO(2))$, which correspond to three complex
structure and three K\"ahler structure moduli. These moduli fields
are always present in symmetric $Z_2\times Z_2$ orbifolds.
The realistic free fermionic models are constructed by adding
additional boundary condition basis vectors, beyond the $Z_2\times Z_2$
twistings. The additional vectors break the $SO(10)$ gauge symmetry down
to a subgroup and reduce the number of generations to three. Their
effect on the untwisted moduli space is extracted by focussing on the 
boundary conditions of the internal world--sheet fermions that correspond
to the six dimensional compactified coordinates. The three generation free
fermionic models give rise to the possibility of assigning asymmetric
boundary conditions to the left and right--movers.
These assignments are reflected in the combinations of
the real internal world--sheet fermions into complex fermions,
or into Ising model world--sheet fermions. The second case 
corresponds to symmetric assignment of boundary conditions,
whereas the first corresponds to asymmetric assignments,
that distinguish between the left-- and right--moving fermions.
This possibility of assigning asymmetric boundary conditions
has important phenomenological consequences. For example, 
for the problem of proton stability and the string doublet--triplet
splitting mechanism \cite{ps}. 

By examining concrete three generation free fermionic models it is noted 
that some models employ boundary conditions that are fully symmetric
\cite{modfix}. 
The moduli space of such quasi--realistic models therefore contains
the three complex and three K\"ahler structure moduli of the original
$Z_2\times Z_2$ orbifold. In these models the internal six dimensional
manifold admit a classical geometrical interpretation. However, 
there also exist quasi--realistic free fermionic models that employ
fully asymmetric boundary conditions. In these models all the six internal
real coordinates have the asymmetric identifications
\begin{equation}
X_L+X_R\rightarrow X_L-X_R
\end{equation}
As a consequence all the geometrical untwisted moduli fields are 
projected out in these models. The additional dimensions
in these compactifications are therefore frozen at the 
enhanced symmetry point. These quasi--realistic string 
vacua therefore do not contain additional classical dimensions,
which are therefore fictitious in these models. Namely the 
extra dimensions exist as organizing principle at some
level in the string partition function, but are not reaslized
physically in the low effective field theory. The situation is
similar to the way in which gauge symmetries are broken
in string theory by Wilson lines. Also in this case the models
contain a GUT gauge symmetry at some level of the string
partition function, which is broken by Wilson lines and is
not an explicit symmetry of the low energy effective field theory.

It is of interest to note that in the quasi--realistic heterotic--string
models discussed here the moduli that arise from the twisted sectors are
projected out as well \cite{modfix}. The reason is that the models
correspond to (2,0) rather than (2,2) compactification. In the (2,2)
models the sectors that complement the 16 representation of SO(10) to
27 of $E_6$, also at the same time produce the twisted moduli. In the (2,0)
models these sectors give rise to vectorial 16 representations of the hidden
SO(16) gauge group and the moduli are projected out together with
the 10+1 representations that are embedded in the 27 of E6. 
It should, however, be emphasized that the models may contain additional
moduli. Additional moduli may arise from flat directions of the 
superpotential and from charged moduli. What is noted here is that the moduli
that are identified as coefficients of exactly marginal operators, and are
therefore interpreted as geometrical moduli, are projected out from
the massless spectrum. Hence the geometrical coordinates in these models
are frozen at the enhanced symmetry point.
In these models there is no apparent classical geometry that underlies 
the additional degrees of freedom that are required to restore
the world--sheet reparameterisation invariance.

In the three generation free fermionic models with the fully asymmetric
identification all the extra dimensions are frozen at the maximally enhanced
symmetry point, which up to a rotation is the same as the self--dual
point under T--duality \cite{tduality}. The attractive phenomenological 
structure of these models and the relation between the maximally enhanced
symmetry point and the self--dual point under T--duality raises the
intriguing possibility that the self--duality criteria is pivotal 
to the vacuum selection. 

\section{Phase--space self--duality and trivial selection}

To illustrate further this possibility I discuss the association of 
a self--dual state with a ``vacuum'' state in a completely unrelated
mathematical setting.
Duality and self--duality also play a key role in the recent
formulation of quantum mechanics from an equivalence
postulate \cite{fm}.
An important facet of this formalism is the
phase--space duality, which is manifested due to the
involutive nature of the Legendre transformation.
In the Hamilton--Jacobi formalism of classical mechanics
the phase--space variables are related by Hamilton's generating
function $p=\partial_q{S}_0(q)$.
One then obtains the dual Legendre transformations \cite{fm},
$$
{ S}_0=p\partial_p{ T}_0-{ T}_0~
$$
and
$$
{ T}_0=q\partial_q{ S}_0-{S}_0,~
$$
where ${ T}_0(p)$ is a new generating function
defined by $q=\partial_p{ T}_0$.
Because of the undefinability of the Legendre 
transformation for linear functions, {\it i.e.} for
physical systems with ${ S}_0=A q+B$,  the
Legendre duality fails for the free system, and for the
free system with vanishing energy. 
We can associate a second order differential 
equation with each Legendre transformation \cite{fm}. 
There exist therefore a set of solutions, labelled by $pq=const$,
which are simultaneous
solutions of the two sets of differential equations. These are the 
self dual states under the phase--space duality. 


The Legendre phase--space
duality and its breakdown for the free system are intimately
related to the equivalence postulate, which states
that all physical systems labelled by the function
${ W}(q)=V(q)-E$, can be connected by a coordinate
transformation, $q^a\rightarrow q^b=q^b(q^a)$, defined
by ${ S}_0^b(q^b)={ S}_0^a(q^a)$.
This postulate implies that there
always exist a coordinate transformation connecting  
any state to the state ${ W}^0(q^0)=0$. Inversely, this means
that any physical state can be reached from the
state ${ W}^0(q^0)$ by a coordinate transformation.
This postulate cannot be consistent
with classical mechanics. The reason being that in Classical
Mechanics (CM) the state ${ W}^0(q^0)\equiv0$ remains a fixed
point under coordinate transformations. Thus, in CM it
is not possible to generate all states by a coordinate
transformation from the trivial state. From
the Classical Hamilton--Jacobi Equation (CHJE) it is seen that
 ${ S}_0=A q+B$ is the solution associated with $V(q)=0 ~\& ~E=const$, 
that is the state for which the Legendre duality breaks down. 
Consistency of the
equivalence postulate therefore implies that $S_0(q)$ is
not a solution of the CHJE, but rather 
a solution of the Quantum Stationary
Hamilton--Jacobi Equation (QSHJE),
$$
({1/{2m}})\left({{\partial_q S}_0}\right)^2+
V(q)-E+({\hbar^2/{4m}})\{{ S}_0,q\}=0,~
$$
where $\{,\}$ denotes the Schwarzian derivative.
The remarkable property of the QSHJE, which distinguishes
it from the classical case, is that it admits a non--trivial solution
also for the trivial state, ${ W}(q)\equiv0$.
In fact the QSHJE implies that ${ S}_0=constant$ is
not an allowed solution. The fundamental characteristic
of quantum mechanics in this approach is that ${ S}_0\ne Aq+B$.
Rather, the solution for the trivial state, with $V(q)=0$ and $E=0$,
is given by
$$
{ S}_0=i\hbar/2\ln q,~
$$
up to M\"obius transformations. Remarkably, this quantum
trivial state solution coincides with the self--dual state
of the Legendre phase--space transformation and its dual.
We have that the quantum self--dual state plays a pivotal
role in ensuring both the consistency of the equivalence
postulate and definability of the Legendre phase--space  
duality for all physical states. 
Furthermore, it is noted that the self--dual state under phase--space duality 
is associated with the state with $V(q)=0$ and
$E=0$. Hence providing
another hint at the association between self--duality and
trivial states in the space of all allowed states. 

\section{Conclusions}

Existence of quasi--realistic string vacua in which all
the untwisted and twisted sectors moduli are projected out was 
demonstrated. In such models the extra dimensions are fictitious.
This may indicate that extra dimensions are fictitious in
phenomenologically viable string vacua. This is an appealing
proposition. While string theory requires additional degrees
of freedom, beyond the four space--time, the interpretation
of those as extra physical dimensions is naive. Extra dimensions
provide an organizing principle for the string symmetries,
but are not realized as physical dimensions in the low 
energy effective field theory. It is the
intrinsic left--right independence of the closed string modes, which 
allows for asymmetric boundary conditions, and results in the
projection of all the K\"ahler and complex structure moduli.
Thus, string theory, which needs the extra degrees of freedom
for its consistency, also provides the intrinsic mechanism
to fix the moduli. The mechanism afforded utilises the
quantum nature of the extra dimensions, and therefore
may indicate the limitation of the effective field theory
analysis. It may also point to the possibility that dilaton
fixation may have to await the quantum formulation of M--theory. 

It is proposed further that phase--space duality is the guiding 
property in trying to formulate quantum gravity. In this
respect T--duality is a key property of string theory.
We can think of T-duality as a phase--space duality
in the sense that we are exchanging momenta and winding
modes in compact space. We can turn the table around
and say that the key feature of string theory is that
it preserves the phase--space duality in the compact
space. Namely, prior to compactification the wave--function 
of a point particle $\Psi\sim{\rm Exp}(i P X)$ is invariant under
$p\leftrightarrow x$. However, in the ordinary Kaluza--Klein
compactification this invariance is lost due to the 
quantization of the momentum modes. String theory restores
this invariance by introducing the winding modes.
It is further argued that the self--dual points
under phase--space duality are intimately
connected to the choice of the vacuum. 

\section*{Acknowledgements}

I would like to thank the Oxford Theoretical Physics Department for
hospitality.
This work is supported in part by the PPARC and by the Royal Society.

\end{document}